\documentclass[12pt]{article}
\usepackage{graphicx,verbatim,array,multicol,palatino,amsfonts,amsmath,subfigure, psfrag}
\usepackage{chicago}
\usepackage{color}

\def\lboxit#1{\vbox{\hrule\hbox{\vrule\kern6pt
      \vbox{\kern6pt#1\kern6pt}\kern6pt\vrule}\hrule}}

\def\thick#1{\hbox{\rlap{$#1$}\kern0.25pt\rlap{$#1$}\kern0.25pt$#1$}}

\def\bfepsilon{{\bf \epsilon}}

\begin{document}

\title{On Bayesian Curve Fitting Via Auxiliary Variables}
\author{Y. Fan\\
{\small School of Mathematics and Statistics}\\
{\small University of New South Wales, Sydney 2052, Australia}\\
\and
J.-L. Dortet-Bernadet \\
{\small Institut de Recherche Math\'ematique Avanc\'ee, UMR 7501 CNRS}\\
{\small Universit\'e de Strasbourg, Strasbourg, France} \\
\and
S. A. Sisson\\
{\small School of Mathematics and Statistics}\\
{\small University of New South Wales, Sydney 2052, Australia}
}

\maketitle


\abstract

In this article we revisit the auxiliary variable method
introduced in \shortciteN{smith+kohn96} for the 
fitting of $P$-th order spline regression models with an unknown number of knot points. We 
introduce modifications which  allow the location of knot points to be random, and we further consider an extension of the method to handle models with non-Gaussian errors. We provide a new algorithm for the MCMC sampling 
of such models. Simulated data examples are used to compare the performance of our method with 
existing ones.  Finally, we
make a connection with some change-point problems, and show how they can be re-parameterised to the variable selection setting.
\\
Supplemental materials including R computing codes used in the examples are available online.


\vskip 2cm

\section{Introduction}
\label{sec:intro}

This article examines methods for Bayesian curve fitting. Specifically, given observation pairs $(x_1, y_1), \ldots, (x_n, y_n)$, we are interested in fitting the regression model
\begin{equation}\label{eqn:linmod}
Y_i|x_1,...,x_n \sim f(x_i)+\epsilon_i, \quad i=1,\ldots,n
\end{equation}
where $\epsilon_1,...,\epsilon_n$ are independent draws from a Gaussian distribution $N(0,\sigma^2)$ with $\sigma > 0$ unknown. The curve $f$, about which we wish to make inference, is a smooth real-valued function defined on some interval $[a,b]$. Later in this article, we consider the case  where the Gaussian error assumption is relaxed. 

A general and powerful non-parametric approach to the fitting of the curve $f$, is via spline functions of a given  degree, $P \geq 1$. In this setting, $f$ can be written as the linear combination
\begin{equation}\label{eqn:spline}
f(x)=\alpha_0 + \sum_{j=1}^P \alpha_jx^j + \sum_{k=1}^K \eta_k(x-\gamma_k)_+^P, \quad x \in [a, b]
\end{equation}
where $z_+=\max\{0,z\}$ and $\gamma_k, k=1,\ldots,K $ represent the locations of  $K$ knot points (see \shortciteNP{hastie+t90}). Typically,  the degree $P$ is set to equal 3, as  cubic splines are known to approximate locally smooth function arbitrarily well. Under the representation (\ref{eqn:spline}), fitting the curve consists of estimating the number of knots $K$, the knot locations $\gamma_k, k=1,\ldots,K$, and the corresponding regression coefficients $\alpha_j$, $j=0,\ldots ,P$ and $\eta_k$, $k=1,\ldots,K$.  See \shortciteN{ruppertwc03} for an accessible exposition on non-parametric regression models using splines.

Several authors provide methods for Bayesian inference on this model. One such method is to introduce a large number of potential knots, each with a fixed location, from which a significant subset can be selected ({\it e.g.} \shortciteNP{fried+silv}). If $\gamma_k$, $k=1,...,K_{max}$ represent  $K_{max}$ known potential knots,  model (\ref{eqn:linmod}) can be written as the linear model
\begin{equation}\label{eqn:varsele}
Y=X_{\gamma}\beta + \bfepsilon
\end{equation}
where $Y=(y_1,\ldots,y_n)'$, $\beta=(\alpha_0,\alpha_1,\ldots,\alpha_P,\eta_1,\ldots, \eta_{K_{max}})'$, $\bfepsilon=(\epsilon_1, \ldots, \epsilon_n)'$, with design matrix 
$$
X_{\gamma}=({\bf 1},{\bf x}, \ldots ,{\bf x}^P, ({\bf x}-{\bf 1}\gamma_1)^P_+, \ldots , ({\bf x}-{\bf 1}\gamma_{K_{max}})^P_+)
$$ 
where ${\bf x}=(x_1,\ldots ,x_n)'$, and ${\bf 1}=(1,\ldots,1)'$. \shortciteN{smith+kohn96} recognised that a Bayesian variable selection technique ({\it e.g.} \shortciteNP{george+M93}) can be used to carry out inference on the curve. The variable selection they proposed only requires a Gibbs sampler (\shortciteNP{gelfand+S90}), thus the  curve fitting procedure is relatively straightforward and easy to compute. For the selection of the best potential knots,  \shortciteN{denison+MS98} proposed to  use a reversible jump Markov chain Monte Carlo (MCMC) algorithm (\shortciteNP{green95}, \shortciteNP{sisson04}); their method avoids computation of the spline coefficients $\eta_k$ by substituting their least squares estimates. \shortciteN{biller00} provides an alternative reversible jump MCMC algorithm extending to the case for non-Gaussian errors.

All these methods are very efficient in practice for many types of applications. Nevertheless, in some cases, the need to define the discrete set of candidate knots can become a limitation. A  common procedure is to use some of the sorted distinct values of the $x_i$'s as potential knots; for example, \shortciteN{smith+kohn96} recommend placing a potential knot each three to five sorted $x_i$ values when using cubic splines. Clearly, this can be problematic when the $x_i$'s are non-regularly spaced. A solution consists of the placement of knots from a continuous proposal, as in   \shortciteN{dimatteo+gk01}, who extended the approach taken by \shortciteN{denison+MS98}  to a fully Bayesian treatment. They used conjugate priors for the regression parameters and integrated them out of the posterior, and proposed a reversible jump MCMC sampler that runs only on the number and locations of the knots.

In this article we consider a generalization of the auxiliary variable method, first introduced in similar context by \shortciteN{smith+kohn96},  that allows the location of the potential knots to be unknown. This generalization is based on the introduction into the model of intervals in which the potential knots may lie. The proposed method is expected to offer a better fit of the curve to the data, since we consider knots from a continuous space, while retaining the simplicity of a Metropolis-within-Gibbs sampler for inference on  the model. More precisely, we give in Section \ref{sec2} the general set up for our modelling strategy and discuss how inference is carried out. Section \ref{sec3} extends the auxiliary variable modelling approach to the more general setting where we have non-Gaussian errors, and suggests a new algorithm for  MCMC sampling. In Section \ref{sec4} we revisit some change-point detection problems and see that the use of an  auxiliary variable setting is  beneficial  from a  computational point of view. Finally, we conclude with some discussion in Section \ref{sec5}.

\section{Curve fitting via an auxiliary variable approach}
\label{sec2}

\subsection{The model and prior assumptions}

We adopt an auxiliary variable approach by introducing a vector of binary
indicator variables $z_k, k=1,\ldots, K_{max}$, 
$$z_k =\left\{\begin{array}{ll}
1 &\quad \mbox{if there is a knot point } \gamma_k \mbox{ in the interval } 
I_k \mbox{ and } \eta_k\neq 0\\
0 &\quad \mbox{if there is no knot point in the interval }I_k \mbox{ and } \eta_k = 0
\end{array}\right.$$
where $\eta_k$ denotes the spline coefficients  in model (\ref{eqn:varsele}), and the intervals $I_k$ are defined on the range of the $x_i$'s. Each interval $I_k$  contains at most one knot with unknown location $\gamma_k$. In practice, such intervals can be defined either using prior information on regions where a knot is suspected or, in the absence of such prior information, an equal partition of the range may be adopted. We denote the vector $(\gamma_1, \ldots , \gamma_{K_{max}})'$  by $\gamma$  and  consider the product of uniform distributions on the interval as the prior distribution on $\gamma$.

Each possible value for  $\gamma$ gives a model of the form (\ref{eqn:varsele}). Let $X_{z, \gamma}$ denote the matrix constructed with the columns of $X_{\gamma}$ corresponding to non-zero entries in $z$, and let $\beta_{z,\gamma}$ denote the vector of corresponding regression coefficients.  We use the following decomposition of the joint prior distribution of all the unknown parameters
$$
\pi( \beta_{z,\gamma},z, \sigma^2, \gamma )=\pi_{\beta_{z,\gamma}}(\beta_{z,\gamma}|z, \sigma^2, \gamma)\pi_{\sigma^2}(\sigma^2)\pi_z(z)\pi_{\gamma}(\gamma),
$$
where
\begin{equation}\label{unitinfo}
\pi_{\beta_{z,\gamma}}(\beta_{z,\gamma}|z, \sigma^2, \gamma) = N(0,\sigma^2c(X'_{z,\gamma}X_{z,\gamma})^{-1}). 
\end{equation}
This conditional prior for $\beta_{z,\gamma}$, related to $g$-priors (\shortciteNP{zellner86}), has the advantage of conjugacy when $\epsilon$ is Gaussian,   in which case the regression and variance parameters can be analytically integrated out. 
The case $c=n$ corresponds to the unit information prior used by \shortciteN{dimatteo+gk01}, a default choice that has worked well in practice with large sample sizes. \shortciteN{smith+kohn96} recommend values of $c$ in the range $10 \leq c \leq 1000$. For the variance parameter, we use the classical uninformative prior $\pi_{\sigma^2}(\sigma^2) \propto 1/\sigma^2$ that  leads to proper  posteriors here (see for example \shortciteNP{gelmancsr03}, Chapter 2). 
Finally, we need to define the prior distribution for $z$. We consider here the decomposition of this prior given by
$$
\pi_z(z)=\pi(z\mid |z|)\pi(|z|) 
$$
where $|z|=\sum_{k=1}^{K_{max}}z_k$ is the number of non-zero entries in $z$, {\it i.e.} the number of knots that are used in the corresponding model. We use as prior for $|z|$
a right-truncated Poisson distribution with parameter $\lambda$, and maximum value $L$.
The value of $L \leq K_{max}$ corresponds to the maximum number of knots allowed. 
We assume  that, given the quantity $L$, all possible configurations for $z$ have equal probabilities, so that
\begin{equation}\label{numprior}
\pi_z(z) \propto
\frac{\lambda^{|z|}}{|z|!}{\boldsymbol 1}_{\{|z|\leq L\}},
\end{equation}
\\
where ${\boldsymbol 1}_{\{A\}}$ is 1 if $A$ is true and 0 otherwise. Under these prior and Gaussian error assumptions, the parameters $\beta_{z,\gamma}$ and $\sigma^2$ are easily integrated out of the posterior distribution. We finally get the joint posterior distribution for $(z,\gamma)$ of the form
\begin{equation}\label{eqn:postconj}
\pi(z, \gamma|Y)\propto (c+1)^{-(|z|+P+1)/2}S_{z,\gamma}(Y)^{-n/2}\pi_z(z)\pi_{\gamma}(\gamma)
\end{equation}
where 
$$
S_{z,\gamma}(Y)=Y'Y-\frac{c}{c+1}Y'X_{z,\gamma}(X'_{z,\gamma}X_{z,\gamma})^{-1}X'_{z,\gamma}Y.
$$

\subsection{Inference on the posterior distribution}
\label{sec:mcmc}

An MCMC sampler is used for the inference on the model. Based on the posterior distribution (\ref{eqn:postconj}), it uses the following successive updates for $z$ and $\gamma$:
\begin{itemize}
\item {\bf Update $z$}. This update involves two types of moves; with probability 0.5 we propose
an add/delete step, otherwise a swap step is proposed. 
Specifically, the two move steps involve
\begin{itemize}
\item add/delete: randomly select a $z_k$ and propose to change its value;
\item swap: randomly select two values $z_i$ and $z_j$, 
and propose to exchange their values. 
\end{itemize}
In both cases, proposed moves from current value $z$ to proposed value 
$z'$ are accepted with the usual Metropolis-Hastings acceptance probability
$$
\alpha(z, z') = \mbox{min} \left\{1, \frac{\pi(z', \gamma|Y) }{\pi(z, \gamma|Y) } \right\} .$$

\item {\bf Update $\gamma$}. For each $k=1,\ldots, K_{max}$, we differentiate the cases when $z_k=0$ and when $z_k=1$:  
\begin{itemize}
\item if $z_k=0$ then $\gamma_k$ is updated according to its prior distribution, {\it i.e.} a  uniform distribution on $I_k$; 
\item  
if $z_k=1$, $\gamma_k$ is updated to a new value $\gamma'_k$, according to the posterior distribution
$$\pi_(\gamma_k|\gamma_{j\neq k},z,Y)\propto S_{z,\gamma}(Y)^{-n/2}\pi_{\gamma}(\gamma).$$
\end{itemize}
An  independence Metropolis-Hastings step can be used for this last type of updating, using the prior on $\gamma_k$ as a proposal, with the corresponding acceptance probability given by
$$
\alpha(\gamma_k, \gamma'_k) = \mbox{min} \left\{1, \frac{\pi(\gamma'_k|\gamma'_{j\neq k},z,Y)}{\pi(\gamma_k|\gamma_{j\neq k},z,Y)}
\right\} .$$
\end{itemize}
%
%
Note that a rejection sampler may alternatively be used for this step, again using the prior on $\gamma_k$
as the sampling distribution. This may be desirable in certain circumstances as the rejection sampler produces i.i.d. draws from the conditional posterior distribution of $\gamma_k$.

Once an MCMC sample $\{(z^{(i)},\gamma^{(i)})\}_{i=1,...,N}$ is obtained, model inference proceeds following one of the two methods commonly used in such settings. The first method uses the  maximum a posteriori (MAP) estimate for $(z,\gamma)$
$$ 
(\hat{z}, \hat{\gamma}) = \underset{1\leq i \leq N}{\mbox{argmax }} \pi(z^{(i)}, \gamma^{(i)} |Y),
$$
and then  calculate the corresponding least squares estimates $\hat{\beta}_{\hat{z},\hat{\gamma}}$ to give the curve estimate
\begin{equation}\label{eqn:MAP}
\hat{f}(x) =X_{\hat{z}, \hat{\gamma}}\hat{\beta}_{\hat{z},\hat{\gamma}} .
\end{equation}
The second method uses a Bayesian model averaging approach  (BMA) where the estimates 
for $\hat{f}(x)$ are averaged over different configurations of the auxiliary variable $z$ and their corresponding $\gamma$ values from the MCMC output. Since the conditional posterior expectation for  $\beta$ given $z$ and $\gamma$ is, for large $c$,
$$
E(\beta_{z,\gamma}|z,\gamma, Y) =\frac{c}{c+1}(X'_{z,\gamma}X_{z,\gamma})^{-1}X'_{z,\gamma}Y \approx \hat{\beta}_{z,\gamma},
$$
where $\hat{\beta}_{z,\gamma}$ is the least squares estimate for $\beta$ given $z$ and $\gamma$, then an estimate for the curve can be obtained by 
\begin{equation}\label{eqn:BMA}
\hat{f}(x)=\frac{1}{N}\sum_{i=1}^N X_{z_i,\gamma_i}\hat{\beta}_{z_i,\gamma_i}.
\end{equation}

\vskip 0.7cm

\subsection{Simulation studies}

We carry out simulation studies using the examples from  \shortciteN{smith+kohn96},
\shortciteN{dimatteo+gk01} and \shortciteN{denison+MS98}. We compare the performance
of the methods of \shortciteN{smith+kohn96} and \shortciteN{denison+MS98} with our proposed method, and also discuss the selection of intervals $I_k$. In 
each example 
a cubic spline model is fitted by setting $P=3$ in 
(\ref{eqn:spline}).  \\

{\it Example 1:} In this example, taken from \shortciteN{smith+kohn96}, the true function takes
the form 
$$f(x) = \phi(x, 0.15, 0.05^2)/4 + \phi(x, 0.6, 0.2^2)/4, \quad x\in [0,1]$$
where $\phi(x,\mu,\sigma^2)$ denotes the value at $x$ of the normal density with mean $\mu$ and variance $\sigma^2$. Some $n$ data points $x$ are sampled from the uniform distribution 
$U(0,1)$, and a zero-mean Gaussian noise $\epsilon$ is added to the data, where $\epsilon \sim N(0, 0.25^2)$. Sample sizes of $n=20$ and $n=100$ are studied.\\

{\it Example 2:}  In this example taken from \shortciteN{denison+MS98} the true
function is
$$
f(x) = \sin(2x) + 2\exp(-16x^2),  \quad x \in [-2,2].
$$
This function is first rescaled so that the support is on the unit interval, and then evaluated at $n$
points in $[0,1]$, generated from a $U(0,1)$ distribution. A zero-mean Gaussian noise $\epsilon$ is then added to the data, where $\epsilon \sim N(0, 0.3^2)$. Sample sizes of $n=20$ and $n=200$ are studied.\\

{\it Example 3:} This example is taken from \shortciteN{dimatteo+gk01}. The true function is
$$f(x) = \sin(x) + 2\exp(-30x^2),  \quad x \in [-2,2],$$
evaluated at $n$ regularly spaced grid points, and the variance of the noise is taken
as $\sigma^2=0.3^2$. Again, we rescale to work on the unit interval for $x$. Sample sizes of $n=20$ and $n=101$ are studied.
\\

\noindent To compare the different methods we use  the mean squared error (MSE) as a measure of goodness of fit, given by
$$\mbox{MSE}=\frac{1}{n}\sum_{i=1}^n \{\hat{f}(x_i)-f(x_i)\}^2$$
where $f$ is the true function and $\hat{f}$ is the estimated function. For each example and for the three methods that are considered, MSEs for maximum a posteriori estimates and  Bayesian model averaging estimates were calculated using Equations (\ref{eqn:MAP}) and (\ref{eqn:BMA}). Hereafter we refer to the data sizes of $n=100, 200, 101$ as large data sets, and $n=20$ as small data sets.

Concerning prior specifications, for each example the value $c=n$ was used for the  prior  (\ref{unitinfo}) when
computing for the larger data sets, and $c=200$ for smaller data sets. As stated in \shortciteN{smith+kohn96}, the value of $c$ should be between 10 to 1000, and in general we found values of around 100 to 500 to give very stable results. For the truncated 
Poisson prior (\ref{numprior}) we set $\lambda=3$ and $L=10$.  We chose the Poisson parameter $\lambda$ to be  3 in the examples below, but results are largely insensitive to values of $\lambda$ around this range.  The maximum number of knots allowed $L$ is chosen to be large enough to not affect the simulation results here. 

For these examples we consider the situation where there is no prior information on the knot locations and chose the intervals $I_k$  to correspond to the ranges given by every $n_x$ sorted $x$ values.  We found that $n_x=4, 10$ and $4$ respectively were sufficient to provide a good fit in each of the three larger data set examples. For $n=20$  we used $n_x=2$ in all three examples. In this case, the use of a B-spline basis  to formulate the $X_{\gamma}$ matrix, as in \shortciteN{dimatteo+gk01},  is required to avoid numerical instability  (see  {\it e.g.} \shortciteNP{ruppertwc03}). In general, the choice of the size of the interval can depend on 
the data and there is a trade-off between computational time and accuracy, 
as sampler convergence is achieved more quickly for smaller number of intervals.
 
\begin{figure}[htb]\begin{center}
\psfrag{BMA}[l][l]{{\tiny BMA}}
 \psfrag{MAP}[l][l]{{\tiny MAP}}
 \psfrag{true curve}[l][l]{{\tiny true curve}}
 \psfrag{x}[l][l]{{\tiny x}}
\psfrag{y}[l][l]{{\tiny y}}
\psfrag{0.2}[t][t]{{\tiny 0.2}}
\psfrag{0.4}[t][t]{{\tiny 0.4}}
\psfrag{0.6}[t][t]{{\tiny 0.6}}
\psfrag{0.8}[t][t]{{\tiny 0.8}}
\psfrag{0.0}[t][t]{{\tiny 0.0}}
\psfrag{1.0}[t][t]{{\tiny 1.0}}
\psfrag{-2}[t][t]{{\tiny -2}}
\psfrag{-1}[t][t]{{\tiny -1}}
\psfrag{0}[t][t]{{\tiny 0}}
\psfrag{2}[t][t]{{\tiny 2}}
\psfrag{1}[t][t]{{\tiny 1}}
\psfrag{-0.5}[t][t]{{\tiny -0.5}}
\psfrag{0.5}[t][t]{{\tiny 0.5}}
\psfrag{1.5}[t][t]{{\tiny 1.5}}
\psfrag{2.0}[t][t]{{\tiny 2.0}}
\begin{tabular}{cc}
\subfigure[Example 1]
{
\includegraphics[height=7cm, angle=-90]{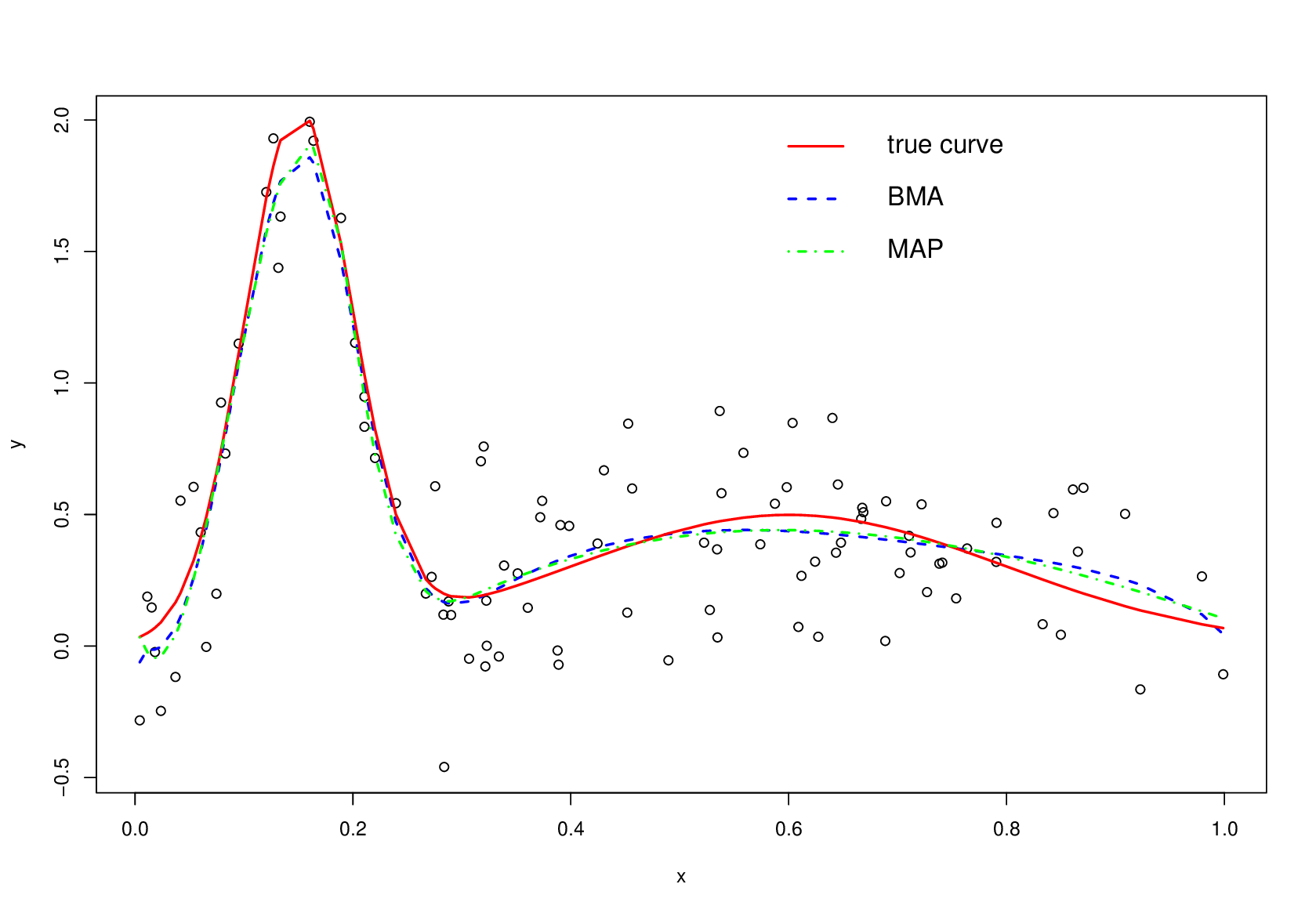}
}
&
\subfigure[Example 2]
{
\includegraphics[height=7cm, angle=-90]{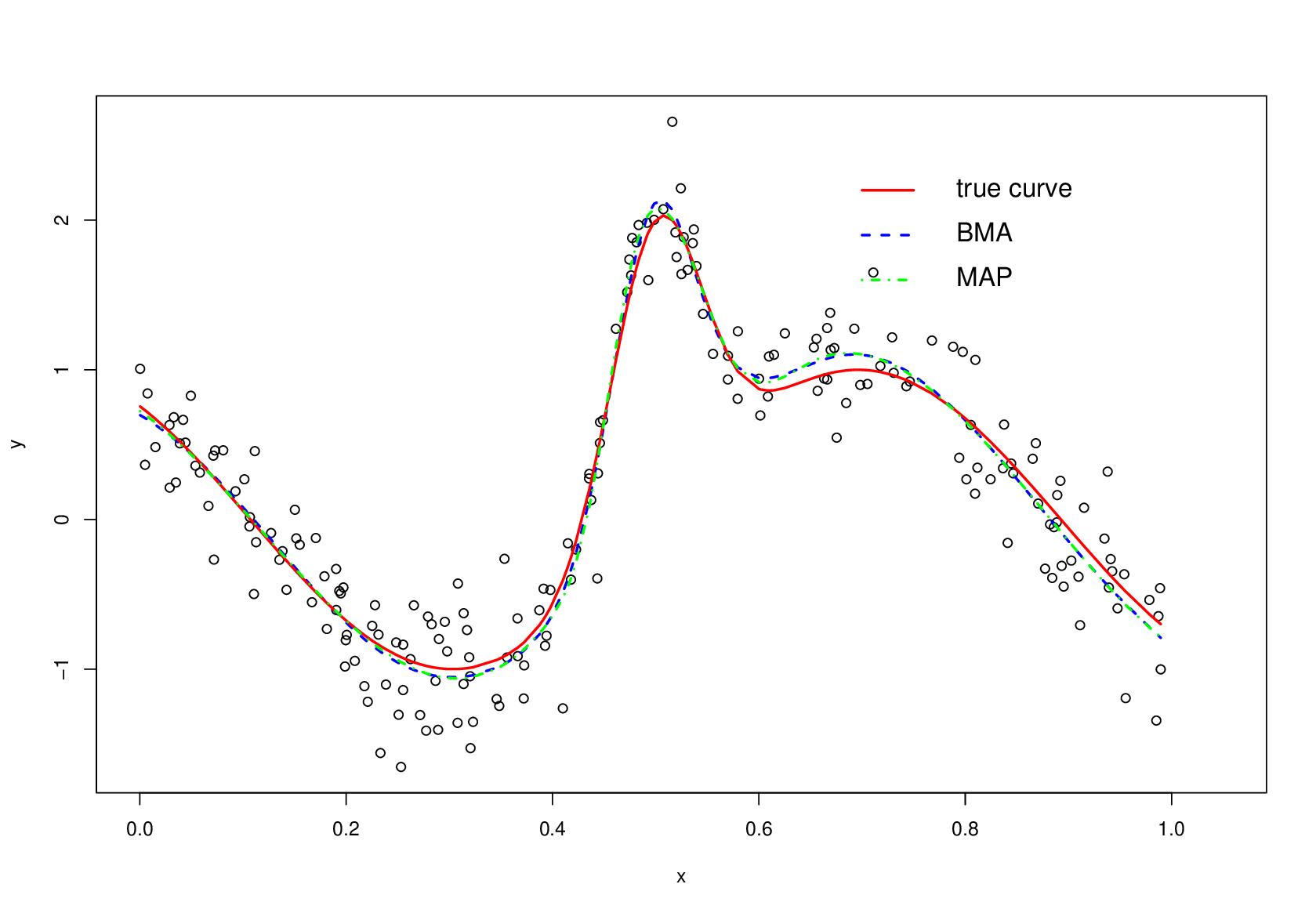}
}
\\
\subfigure[Example 3]
{
\includegraphics[height=7cm, angle=-90]{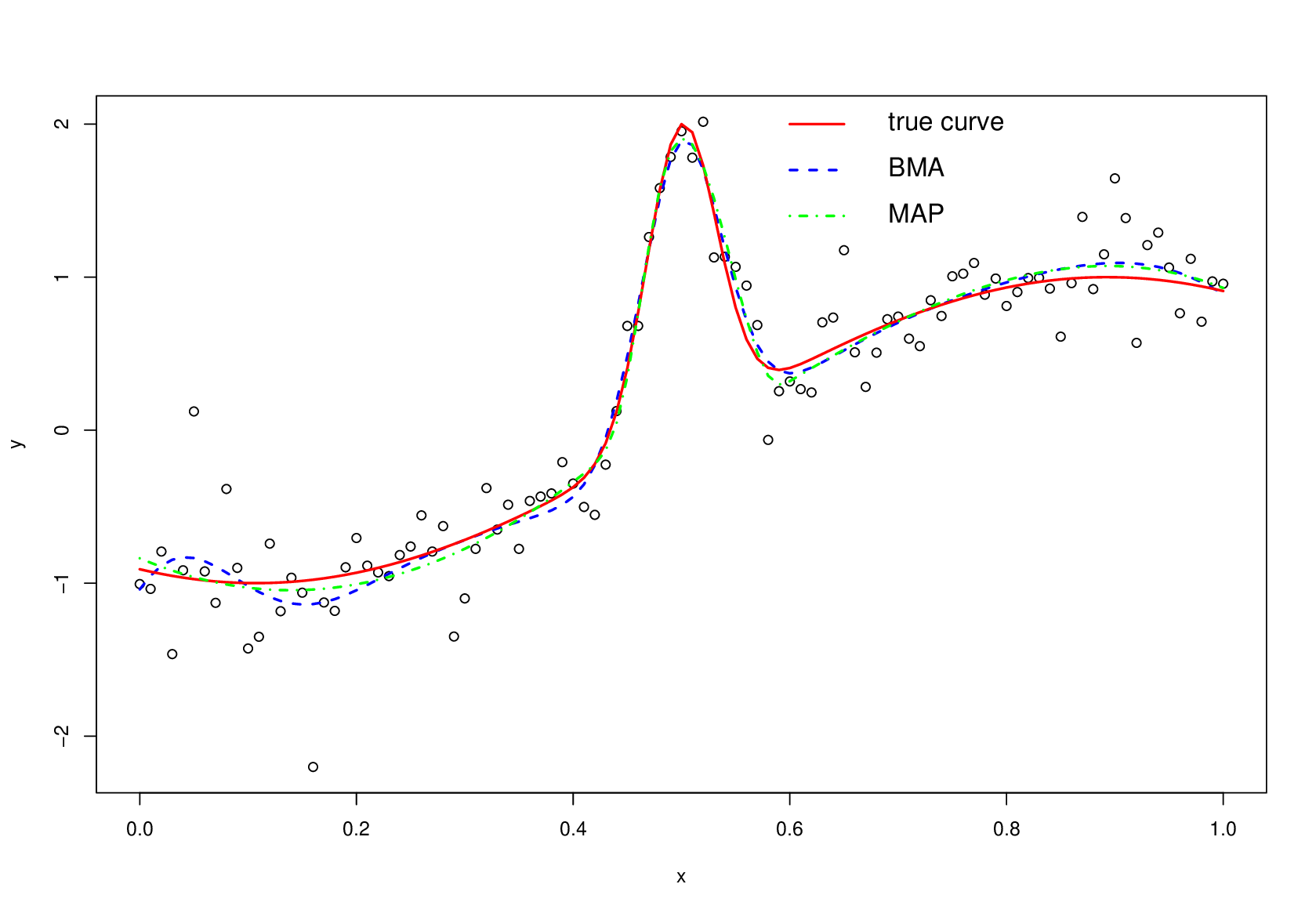}
}
\end{tabular}
\caption{Fitted cubic splines for $n=100, 200, 101$ respectively for the three simulated examples.
The three curves plotted are the true curve (solid line), estimated curves using MAP estimate (dash-dotted lines) and BMA estimate (dashed lines). }
\label{fig:examples}
\end{center}\end{figure}

Finally, for the MCMC computation of all three examples,  starting with an arbitrary set of initial values generated from the prior distributions, we ran a burn-in of 500 iterations, followed by 1,000 recorded iterations, where each iteration involves an update of 20  $z$ update steps for each  $\gamma$ update step.  Note that we found it to be more effective to increase the number of $z$ updates, instead of increasing the total number of iterations,  as $\gamma$ updates had very good mixing properties.
To assess convergence, we monitored the trace plots of posterior values.  We also ran much longer chains of 10,000 iterations and found the  results to be similar in terms of MSE calculations.  This is perhaps not surprising since the posterior values suggested that the chains mixed very quickly. See Figure \ref{fig:examples} for the fitted functions of the three examples using our method.

\begin{figure}[htb]\begin{center}
\psfrag{Ex 1, MAP}[l][l]{{\tiny Ex 1, MAP}}
\psfrag{Ex 2, MAP}[l][l]{{\tiny Ex 2, MAP}}
\psfrag{Ex 3, MAP}[l][l]{{\tiny Ex 3, MAP}}
\psfrag{Ex 1, BMA}[l][l]{{\tiny Ex 1, BMA}}
\psfrag{Ex 2, BMA}[l][l]{{\tiny Ex 2, BMA}}
\psfrag{Ex 3, BMA}[l][l]{{\tiny Ex 3, BMA}}
\psfrag{FDS}[l][l]{{\tiny FDS}}
\psfrag{SK}[l][l]{{\tiny SK}}
\psfrag{DGK}[l][l]{{\tiny DGK}}
\psfrag{0.20}[t][t]{{\tiny 0.20}}
\psfrag{0.15}[t][t]{{\tiny 0.15}}
\psfrag{0.10}[t][t]{{\tiny 0.10}}
\psfrag{0.05}[t][t]{{\tiny 0.05}}
\psfrag{0.00}[t][t]{{\tiny 0.00}}
\psfrag{0.02}[t][t]{{\tiny 0.02}}
\psfrag{0.04}[t][t]{{\tiny 0.04}}
\psfrag{0.06}[t][t]{{\tiny 0.06}}
\psfrag{0.08}[t][t]{{\tiny 0.08}}
\includegraphics[height=\textwidth, angle=-90]{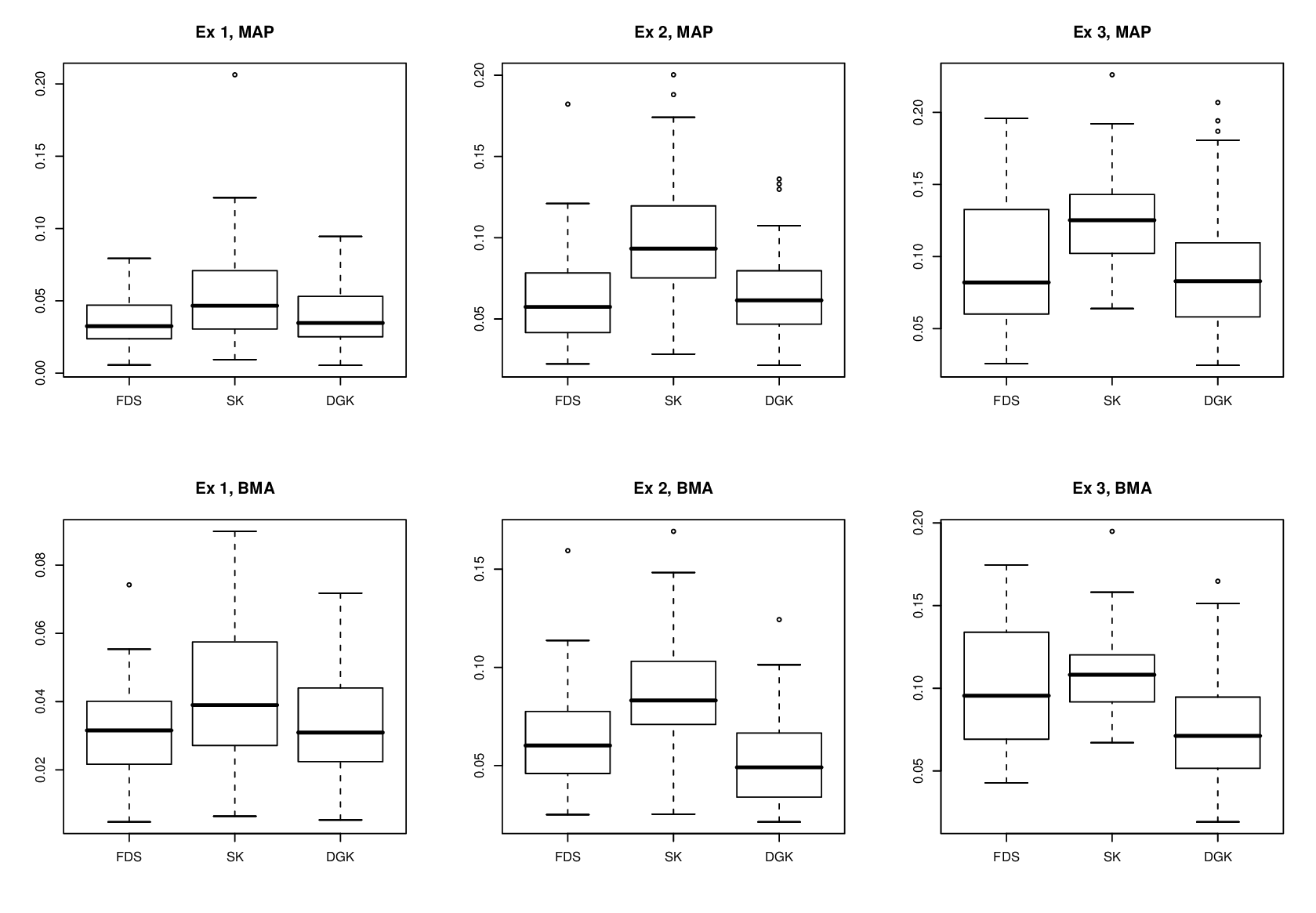}
\caption{Simulation study. Boxplots corresponding to the MSEs presented in Table \ref{table1} for small sample sizes $n=20$. }
\label{fig:boxplots}
\end{center}\end{figure}

Table  \ref{table1} shows the MSEs for both the MAP and the BMA estimates using our method,  the method of \shortciteN{smith+kohn96} (using 1,000 iterations of MCMC updates and 500 burn in) and the method of \shortciteN{dimatteo+gk01} (using 10,000 iterations and 1,000 burn in, as recommended in their paper).  The method of \shortciteN{dimatteo+gk01} was tested using the BARS program available at {\tt http://wpicr.wpic.pitt.edu/WPICCompGen/bars.htm}.  For each example, estimates are calculated over 50 randomly generated data sets, respectively for both large and small data set sizes. See also Figure \ref{fig:boxplots} for boxplots of these MSEs for the small data sets.

In all examples, particularly for the smaller sample size of $n=20$, the method presented in this paper clearly out performed the method of \shortciteN{smith+kohn96} in both the MAP and BMA estimates.  Our method is also very competitive with the method of \shortciteN{dimatteo+gk01}. This is particularly noticeable  in Example 1, where both our MAP and BMA estimates are marginally better, while in Examples 2 and 3 the MAPs generally performed better than BMA when compared to \shortciteN{dimatteo+gk01}. 
The differences between the three methods for larger data sets are smaller, with our method out performing the method of \shortciteN{smith+kohn96} by an order of between $10^{-3}$ to $10^{-4}$ in MSE estimates.  

Overall, our sampler clearly out performed the method of \shortciteN{smith+kohn96}. This gain in  accuracy can be mainly attributed to the fact that our method allows a free knot selection procedure. Our sampler is also more efficient at finding the MAP estimate, resulting in 
smaller MSE estimates than \shortciteN{dimatteo+gk01} in general, while our corresponding BMA estimates are less accurate.
This is perhaps unsurprising, since our algorithm contains more parameters, hence it would be difficult to visit every configuration the appropriate number of times. On the other hand, our algorithm is able to traverse the region of high density very quickly, hence obtaining an accurate MAP estimate in a relatively short number of iterations.

In terms of computation, there are two main differences between our method and that of \shortciteN{smith+kohn96}. Firstly, for the update of $z$, while \shortciteN{smith+kohn96} cycle through each component of the $z$ vector  systematically, we randomly update a number of its  components. For example, in the implementation of Example 1 with $n=100$, we update the $z$ vector 20 times compared to 25 times using  \shortciteN{smith+kohn96}; clearly, the computational gain is greater in Example 2 with $n=200$ when the length of the $z$ vector in \shortciteN{smith+kohn96} is 50. Secondly, we have the additional update of the $\gamma$ parameters. However, this is a quick procedure, since it involves a simple sample from the prior distribution for when there is no knot in the interval, and a Metropolis-Hastings update for when there is a knot, but the numbers in the latter are mostly small. A comparison with the  method of \shortciteN{dimatteo+gk01} is more difficult, since they use a reversible jump scheme. So, in this regard, some users may find it simpler to work with our standard MCMC framework.  We have also found that \shortciteN{dimatteo+gk01} required a much longer Markov chain to acheive convergence, particularly in terms of finding the MAP result, suggesting that there may be mixing issues. 

\begin{table}[htb]
\begin{center}
\begin{tabular}{|l|c|c|c|c|c|}
\hline
&&&FDS &SK &DGK \\
\hline
Example 1 &$n=20$ & MAP &0.0355   &0.0531 &0.0387 \\
&&&(0.0169)   &(0.0351) &(0.0203)\\
&&BMA &0.0317  &0.0433  &0.0335 \\
&& &(0.0135) &(0.0199) &(0.0164)\\
&$n=100$ & MAP &0.0072   &0.0078   &0.0078 \\
&&&(0.0036)   &(0.0038)  &(0.0041)\\
&&BMA &0.0066  &0.0073  &0.0060 \\
&& &(0.0032) & (0.0034) &(0.0030)\\
\hline
Example 2 &$n=20$ & MAP &0.0631   &0.0961&0.0664 \\
&&&(0.0288)   &(0.0379) &(0.0273)\\
&&BMA &0.0638  &0.0837  &0.0534 \\
&& &(0.0289) &(0.0308) &(0.0233)\\
&$n=200$ &MAP &0.0070   &0.0088 &0.0068 \\
&& & (0.0029)  & (0.0057) & (0.0027)\\
&&BMA &0.0061 &0.0076 &0.0057 \\
&& & (0.0022) & (0.0029) &(0.0022)\\
\hline
Example 3 &$n=20$  & MAP &0.0936   &0.1262&0.0916 \\
&&&(0.0468)   &(0.0308) &(0.0439)\\
&&BMA &0.1012  &0.1093  &0.0772 \\
&& &(0.0369) &(0.0244) &(0.0359)\\
&$n=101$ &MAP &0.0123  &0.0134  &0.0116\\
&& &(0.0068) & (0.0069) &(0.0055)\\
&&BMA &0.0116  &0.0133  &0.0099 \\
&& & (0.0061) & (0.0060) & (0.0056)\\
\hline
\end{tabular}
\caption{Simulation study. Mean MSEs with estimated standard errors in brackets based on 50 samples obtained using the maximum a posteriori (MAP)  and Bayesian model averaging (BMA) estimates for each of the three  methods: FDS (method presented in this paper), SK (method of  Smith and Kohn 1996) and DGK (method of DiMatteo et al.  2001).}\label{table1}
\end{center}
\end{table}

\section{Extensions to non-Gaussian error models}

\label{sec3}

When the assumption of normality of $\epsilon_i$ in (\ref{eqn:linmod}) is relaxed we can consider a model of the form
\begin{equation}\label{eqn:gen}
Y_i|x_1,...,x_n \sim p(y|f(x_i),\sigma^*), \quad i=1,...,n,
\end{equation}
where $f(x)$ is still given by (\ref{eqn:spline}) and where $\sigma^*$ denotes a potential nuisance parameter. The methodology used to fit the regression model can be employed in this more general setting, with the exception that we can no longer  integrate out the $\beta_{z,\gamma}$ parameters analytically.

\subsection{Inference on non-Gaussian error models}
\label{sec:non-gauss}

We need to add steps to update the values of the $\beta_{z,\gamma}$ parameters in the MCMC sampler of Section \ref{sec:mcmc}. Here $z_k=0$ corresponds to $\eta_{k}=0$, so we first propose to  update the $z$ and the $\beta_{z,\gamma}$ parameters simultaneously and then propose $\kappa$ further updates of $\beta$ to improve mixing. More precisely we use the following successive updates:
\begin{itemize}
\item {\bf update $z$ and $\beta_{z,\gamma}$} :
\begin{itemize}
\item Propose to update $z$ to $z'$ via either an add/delete or a swap step as in Section \ref{sec:mcmc}.
\item Propose a new value $\beta'_{z',\gamma}$ for the regression coefficients according to an independence Metropolis-Hastings sampler. We  use a multivariate Normal distribution
$N(\hat{\beta}_{z',\gamma}, \delta_z\hat{\Sigma}_{z',\gamma})$ as the proposal $q_{z, z'}$ , where
$\hat{\beta}_{z',\gamma}$  is the MLE estimate of $\beta$ and $\hat{\Sigma}_{z',\gamma}$ is the 
 corresponding covariance matrix  with respect to $z'$ and $\gamma$.
 The move to $(z',\beta'_{z',\gamma})$ is then accepted with probability
$$
\mbox{min}\left\{1, \frac{\pi(\beta'_{z',\gamma}, z',\gamma \mid Y) q_{z',z}(\beta'_{z',\gamma}, \beta_{z,\gamma})  }
{\pi(\beta_{z,\gamma}, z,\gamma \mid Y)q_{z,z'}(\beta_{z,\gamma}, \beta'_{z',\gamma})} \right\},
$$
otherwise, the chain remains at $(z,\beta_{z,\gamma})$.
\end{itemize}

\item {\bf update $\beta_{z,\gamma}$} :
\begin{itemize}
\item If the update $z$ and $\beta_{z,\gamma}$ move above is accepted, then 
perform $\kappa \geq 0$ extra Metropolis-Hastings  updates of $\beta$ using the multivariate Normal distribution $N(\beta_{z,\gamma}, \delta_{\beta}\hat{\Sigma}_{z,\gamma}) $ as the proposal   $q_{\beta, \beta'} $. The moves from $\beta$ to $\beta'$ are accepted with probability
 $$
\mbox{min}\left\{1, \frac{\pi(\beta'_{z,\gamma}, z,\gamma \mid Y) q_{\beta',\beta}(\beta'_{z,\gamma}, \beta_{z,\gamma})  }{\pi(\beta_{z,\gamma}, z,\gamma \mid Y)q_{\beta,\beta'}(\beta_{z,\gamma}, \beta'_{z,\gamma})} \right\}.
$$

\end{itemize}

\item {\bf update $\gamma$}:
\begin{itemize}
\item The corresponding  $\gamma$ update step would remain the same as in Section \ref{sec:mcmc}.
\end{itemize}
\end{itemize}
The values $\delta_z$ and $\delta_{\beta}$ can be tuned to optimise the mixing of the MCMC sampler in the usual way (see   \shortciteNP{robertsr01}). In update $\beta$,  we perform $\kappa\geq 0$ further steps of MCMC moves if the chain has moved to a new model 
to further facilitate mixing of the Markov chain, this step can be omitted for a longer overall MCMC chain. Note that for some applications, the computational cost of estimating the MLE of the likelihood may be similar to estimating the maximum a posteriori estimator of the posterior distribution. In this case we recommend the use of the latter to form the proposal distributions since this give higher acceptance probabilities, see Example \ref{sec:rainfall}.

When the posterior differ greatly from the likelihood, making $q_{z, z'}$ in the update model move a poor proposal choice. In this situation, one may delay the rejection of the move from 
$z$ to $z'$ by making the $\kappa \geq 0$ additional update ($\beta' \rightarrow \beta^*$) moves first with respect to some distribution $\pi^*$,  then carry out the accept/reject decision from $(z, \beta_{z,\gamma})$ to $(z', \beta^*_{z',\gamma})$ with acceptance probability
 $$
\mbox{min}\left\{1, \frac{\pi(\beta^*_{z',\gamma}, z',\gamma \mid Y) \pi^*(\beta'_{z',\gamma}, z', \gamma \mid Y ) q_{\beta',\beta}(\beta'_{z',\gamma}, \beta_{z,\gamma})  }
{\pi(\beta_{z,\gamma}, z,\gamma \mid Y)\pi^*( \beta^*_{z',\gamma}, z', \gamma \mid Y) q_{\beta,\beta'}(\beta_{z,\gamma}, \beta'_{z',\gamma})} \right\}.
$$
Note that such a strategy is only beneficial when the moves $\beta' \rightarrow \beta^*$ are made
with respect to a new distribution $\pi^*$, 
where the distribution $\pi^*$ is chosen to facilitate moves towards the mode of the posterior distribution $\pi$, consequently increasing the acceptance probability in the update model move. See \shortciteN{alhj04} 
for further discussions on how to choose $\pi^*$. For the examples we studied, we did not find it necessary to make use of $\pi^*$, however the reader is referred to \shortciteN{alhj04} should mixing become an issue.

\subsection{A simulated Poisson example}
In this section, we generate $n=500$ Poisson random variables
$y_i, i=1,\ldots, n$
from
$$y_i\sim \mbox{Poisson}(\exp\left\{2 x_{i} +
\cos(4\pi x_{i})\right\})$$
where  $x_{i}$ is uniformly sampled on the interval $[0,1]$.
We fit the curve (\ref{eqn:spline}) for $P=1,2,3$. We take the unit information prior $c=n$ in (\ref{unitinfo})  for the $\beta$ parameter, setting $c=500$, and for the truncated Poisson prior for $z$ we take $\lambda=1$ and $L=10$. We set the intervals $I_k$ to be between consecutive numbers of the sequence
$$(0.02, 0.1, 0.2,.0.3, 0.4, 0.5,0.6,0.7,0.8,0.9,0.98).$$
Note that the first and last interval are bounded away from the limits of the observed points, in order to avoid numerical problems which can sometimes occur with the specification of the prior (\ref{unitinfo}).

\begin{figure}[htb]\begin{center}
\psfrag{true curve}[l][l]{{\tiny true curve}}
\psfrag{P=1}[l][l]{{\tiny P = 1}}
\psfrag{P=2}[l][l]{{\tiny P = 2}}
\psfrag{P=3}[l][l]{{\tiny P = 3}}
\psfrag{x}[l][l]{{\tiny x}}
\psfrag{y}[l][l]{{\tiny y}}
\psfrag{0.2}[t][t]{{\tiny 0.2}}
\psfrag{0.4}[t][t]{{\tiny 0.4}}
\psfrag{0.6}[t][t]{{\tiny 0.6}}
\psfrag{0.8}[t][t]{{\tiny 0.8}}
\psfrag{0.0}[t][t]{{\tiny 0.0}}
\psfrag{1.0}[t][t]{{\tiny 1.0}}
\psfrag{0}[t][t]{{\tiny 0}}
\psfrag{5}[t][t]{{\tiny 5}}
\psfrag{10}[t][t]{{\tiny 10}}
\psfrag{15}[t][t]{{\tiny 15}}
\psfrag{20}[t][t]{{\tiny 20}}
\psfrag{25}[t][t]{{\tiny 25}}
\psfrag{30}[t][t]{{\tiny 30}}
\begin{tabular}{cc}
\subfigure[MAP]
{
\includegraphics[height=8cm, angle=-90]{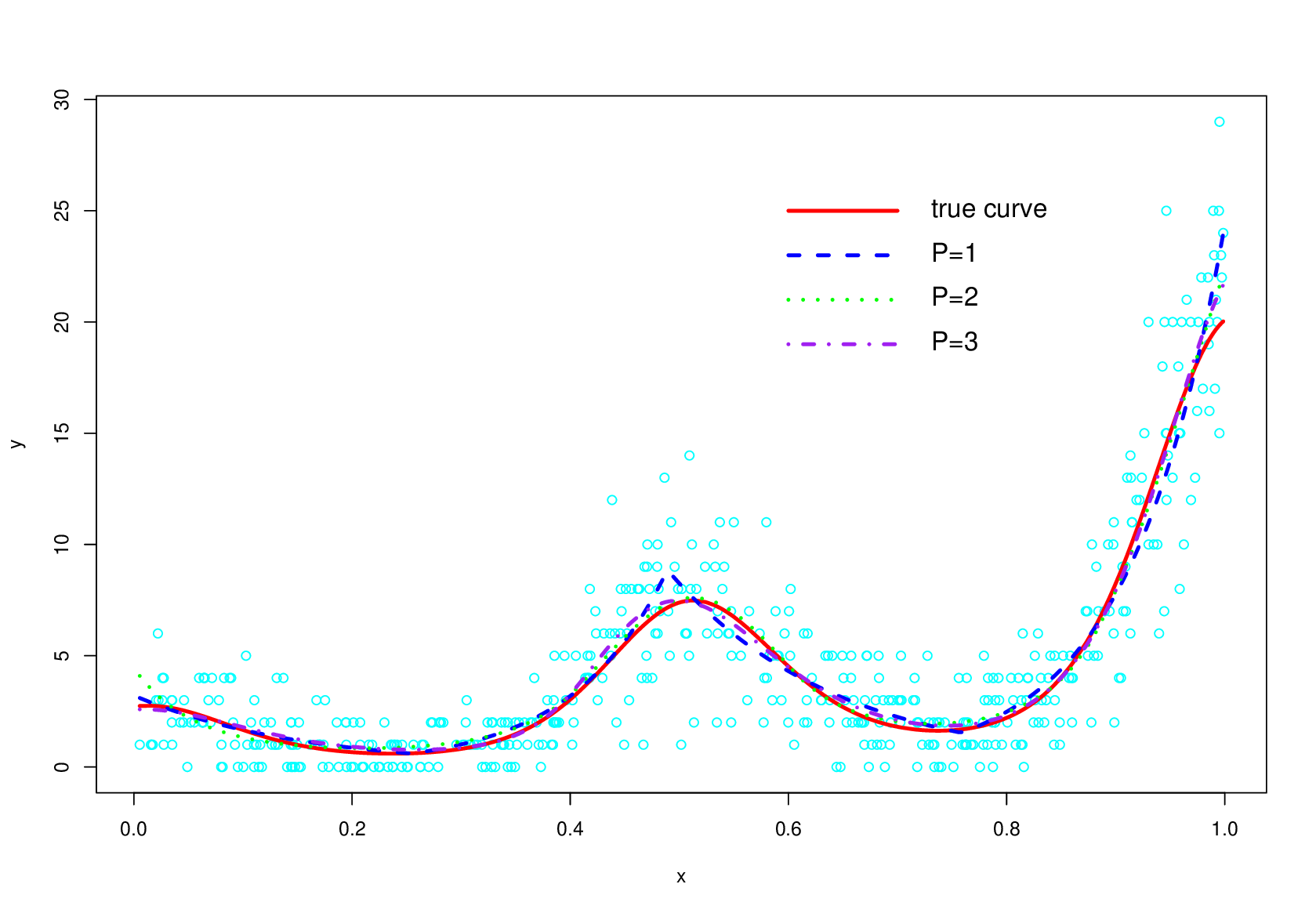}
}
&
\subfigure[BMA]
{
\includegraphics[height=8cm, angle=-90]{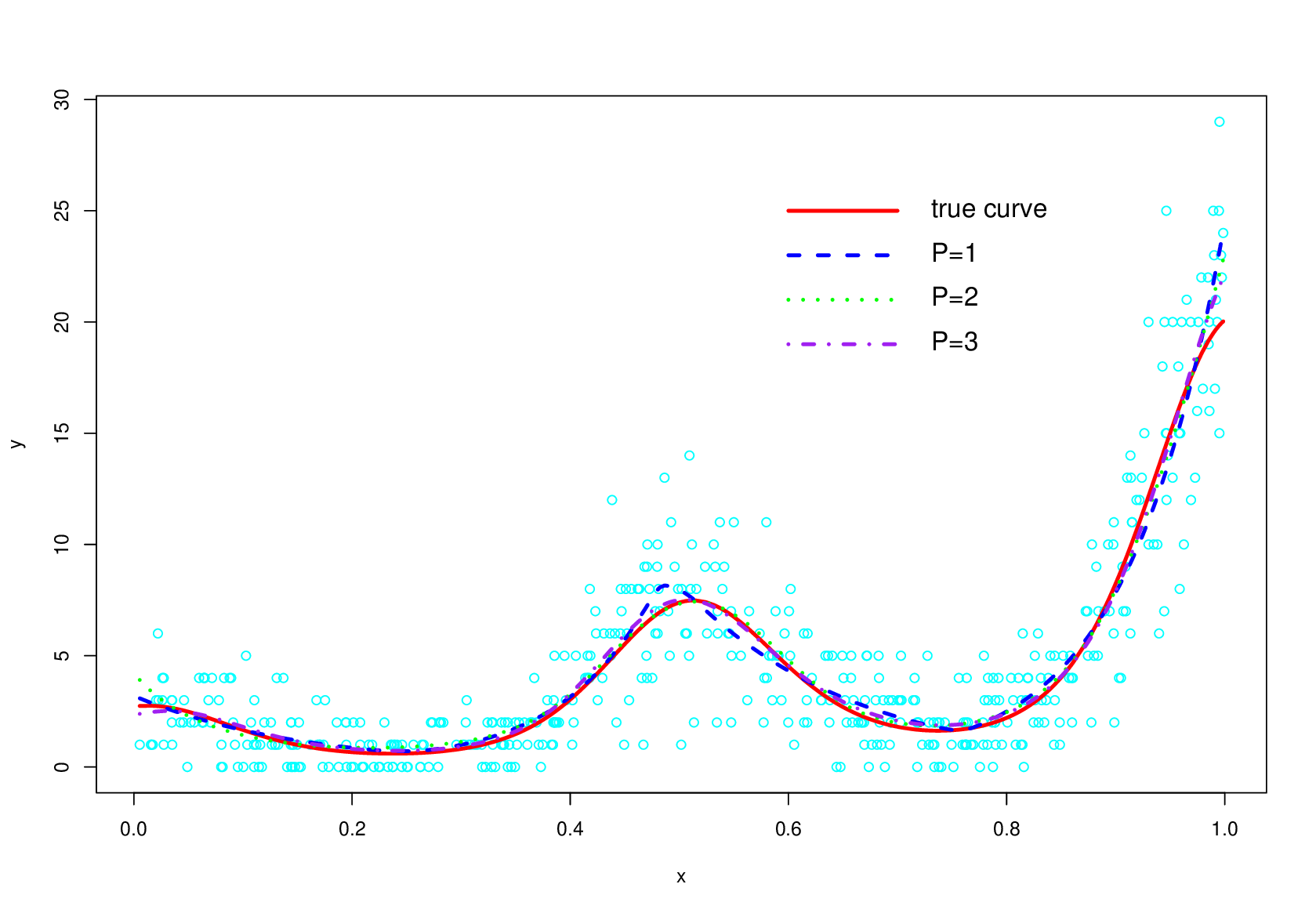}
}
\end{tabular}
\caption{Simulated Poisson example; fitted curves using MAP (left) and BMA (right) estimates for $P=1,2,3$.}
\label{fig:PMM}
\end{center}\end{figure}

Table \ref{tablePMM} shows the average mean squared error calculations obtained from 50 simulated datasets, together with the corresponding standard deviation. For each dataset, we ran our sampler for 5,000 iterations following 1,000 iterations of burn-in. Here, for each update $\gamma$ step, 10 update $z$ and $\beta_{z,\gamma}$ steps are performed to obtain good mixing. For each update of $z$ and $\beta_{z,\gamma}$ we used $\kappa=10$ MCMC moves for $\beta_{z,\gamma}$.  Scaling parameters of the covariance matrices in the proposal distributions for the update of $\beta_{z,\gamma}$ are  $\delta_z=\delta_{\beta}=1/50$. Note that the MAP and the BMA estimates here differ from Equations (\ref{eqn:MAP}) and (\ref{eqn:BMA})  since the values of $\beta_{z,\gamma}$ are not MLE plug-ins. Figure \ref{fig:PMM} shows the fitted curves using the two estimators. The BMA estimates give a smoother curve estimate, particularly for $P=1$.

Two alternative methods of updating the $\beta_{z,\gamma}$ parameters have been used. One is to use the 
MLE plug-in estimates for the $\beta_{z,\gamma}$s as in \shortciteN{denison+MS98},  where $\beta_{z,\gamma}$ is not treated as a parameter in their Bayesian model. In implementing this method for this example, we found it is only slightly quicker than our MCMC update of $\beta_{z,\gamma}$, since for each update of $\beta_{z,\gamma}$ the expense of estimating the MLEs is the same for both algorithms. Our method then includes an additional
$\kappa=10$ computationally inexpensive steps of Metropolis-Hastings updates using the existing MLE estimates.
In an alternative method, \shortciteN{dimatteo+gk01} propose to use an importance sampler to calculate the
expected values of the $\beta_{z,\gamma}$s at each iteration. We implemented this method, using an importance sampling distribution based on the MLE estimates and the corresponding covariance matrix, to obtain 1,000 samples, and found this to be considerably
slower than our method. The MSE estimates using both plug-in MLE and 
importance sampling were approximately the same as found in Table \ref{tablePMM}. 

\begin{table}[htb]
\begin{center}
\begin{tabular}{|l|c|c|c|}
\hline
&$P=1$ &$P=2$ &$P=3$  \\
\hline
 MAP &0.3659 (0.0959)   &0.1647 (0.0725)  &0.1176 (0.0648)\\
 BMA &0.2712 (0.0977) &0.1626 (0.0918) &0.1117 (0.0671)\\
 \hline
\end{tabular}
\caption{Simulation study for the Poisson example. Average  MSEs with estimated standard errors in brackets based on 50 samples, 
obtained using 
maximum a posteriori (MAP) and Bayesian model
averaging (BMA) estimates.}\label{tablePMM}
\end{center}
\end{table}

\section{Applications in change-point modelling}\label{sec4}

Many change-point type problems can be converted to the curve fitting framework. In the following, we first show by example
an explicit equivalence between a  change-point model and a linear regression spline, where one is interested in retaining interpretation
of the coefficients. We then give an example of change-point detection in the context of accurate seasonal modelling for an extreme rainfall problem.

\subsection{Bayesian modelling of prehistoric tombs}

\label{sec:tombs}

Consider the modelling of prehistoric corbelled domes (late Minoan Tholos
data collected from Dimini in Crete; \shortciteNP{cavl82}). 
Paired data arise in the form $(d_i,r_i)$, where $d_i$ represents the distance below
the apex of the tomb with the corresponding radius $r_i$ measured at $d_i$. These data are thought to 
approximately follow a log-linear model between a series of change-points.  The  model  is formulated as
\[ \log(r_i) = 
\log(a_j)+b_j\log(d_i+\Delta_j)+\varepsilon_i, 
\mbox{ if } \gamma_{j-1}\leq d_i<\gamma_j
\]
where $\Delta_j$ is the distance between the apex of the tomb and the begining of the measurement of depth $d_j$. The change-points $0 <   \gamma_1 < \ldots < \gamma_K$ 
(with $K$ unknown) and the parameters of the model are subject to the continuity constraints
\[
	a_j(\gamma_j+\Delta_j)^{b_j} =a_{j+1}(\gamma_j+\Delta_{j+1})^{b_{j+1}}, 
\] 
for $j=1,\ldots,K-1$ and  $\varepsilon_i \overset{i.i.d. }{\sim} N(0, \sigma^2)$. 
We are interested in making posterior inference on the number
and location of the change-points, as well as the coefficients $a_j$ and $b_j$, while retaining their parametric interpretations.

Here, we restrict our interest only to the detection of the number and location of the change points. A more sophisticated model was considered from a Bayesian perspective by \shortciteN{fan+b00}, where computation was
carried out using the reversible jump MCMC algorithm of \shortciteN{green95}, using split/merge and birth/death proposals for the transdimensional moves. The above
representation of a change-point model can be equivalently re-expressed in our framework of
Equation (\ref{eqn:spline}), where the function $f(x)$ is given by
\[
	f(x)=\alpha_0+\alpha_1 x +\sum_{k=1}^K \eta_k(-1+x/\gamma_k)_+,
\]
where $x_i=\log(d_i+\Delta_j)$, $\log(a_1)=\alpha_0$ and $\log(a_j)= \alpha_0 - \sum_{k=1}^{j-1} \eta_k, j>1$ and where $b_1=\alpha_1$ and $b_j= \alpha_1+\sum_{k=1}^{j-1} \eta_k/\gamma_k, j>1$. The corresponding design matrix is given by
\begin{equation}
X_{\gamma}=\left(\begin{array}{llllll}
1 &x_1 &\ldots &(-1+x_1/\gamma_1)_+ &\ldots &(-1+x_1/\gamma_K)_+\\
1 &x_2 &\ldots &(-1+x_2/\gamma_1)_+ &\ldots &(-1+x_2/\gamma_K)_+\\
\ldots &\ldots&\ldots &\ldots&\ldots \\
1 &x_n &\ldots &(-1+x_n/\gamma_1)_+ &\ldots &(-1+x_n/\gamma_K)_+
\end{array}\right).
\end{equation}
This alternative design matrix allows us to retain interpretation on the regression coefficients.
For simplicity, we set the value of $\Delta_j=0.54, j=1,\ldots,K-1$, the value of posterior mean for these parameters found in the model with the highest posterior probability in \shortciteN{fan+b00}. Note that we could incorporate the updating of the $\Delta_j$ parameters into our current algorithm.

\begin{figure}[htb]\begin{center}
\psfrag{posterior}[l][l]{{\tiny posterior}}
\psfrag{iterations}[l][l]{{\tiny iterations}}
\psfrag{No. of knots}[l][l]{{\tiny No. of knots}}
\psfrag{0}[t][t]{{\tiny 0}}
\psfrag{1000}[t][t]{{\tiny 1000}}
\psfrag{2000}[t][t]{{\tiny 2000}}
\psfrag{3000}[t][t]{{\tiny 3000}}
\psfrag{4000}[t][t]{{\tiny 4000}}
\psfrag{5000}[t][t]{{\tiny 5000}}
\psfrag{12}[t][t]{{\tiny 12}}
\psfrag{14}[t][t]{{\tiny 14}}
\psfrag{16}[t][t]{{\tiny 16}}
\psfrag{18}[t][t]{{\tiny 18}}
\psfrag{0.0}[t][t]{{\tiny 0.0}}
\psfrag{0.5}[t][t]{{\tiny 0.5}}
\psfrag{1.0}[t][t]{{\tiny 1.0}}
\psfrag{1.5}[t][t]{{\tiny 1.5}}
\psfrag{2.0}[t][t]{{\tiny 2.0}}
\includegraphics[height=\textwidth,  angle=-90]{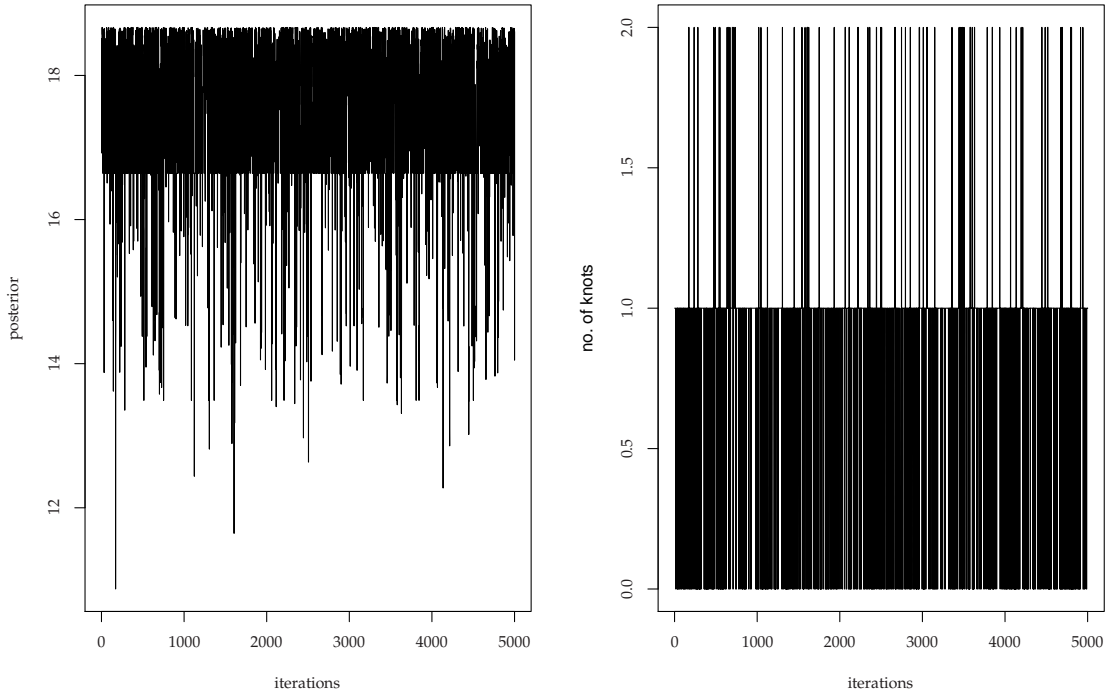}
\caption{Traceplots of the posterior values and the number of change points for the Bayesian model of prehistoric tombs example.}
\label{fig:convMCMC}
\end{center}\end{figure}

Since, in this example, the data consist of only $n=15$ data points, we set the value of $c=500$
in the prior specification of Equation (\ref{unitinfo}) to reflect a vague prior. We take a truncated Poisson prior for the number of change points with $\lambda=1$ and truncated at a maximum
of $L=3$ change points. Visual inspection of the data suggest that it would be sensible to place the interval for the occurrence of change points to be between
the values $0.15, 0.8, 1.2, 1.6$. We ran the MCMC sampler of Section \ref{sec:mcmc} with 1,000 iterations of burn-in and 5,000 iterations of post burn-in samples. To increase mixing, for each update of the $\gamma$ parameter, we updated the auxiliary variable $z$ 20 times. Trace plots of the posterior values and number of knots over the iterations are shown in Figure \ref{fig:convMCMC}. Convergence appears to have been achieved after around 1,000 iterations in this example.

\begin{figure}[htb]\begin{center}
\psfrag{0.5}[t][t]{{\tiny 0.5}}
\psfrag{0.0}[t][t]{{\tiny 0.0}}
\psfrag{0.2}[t][t]{{\tiny 0.2}}
\psfrag{0.4}[t][t]{{\tiny 0.4}}
\psfrag{0.6}[t][t]{{\tiny 0.6}}
\psfrag{1.2}[t][t]{{\tiny 1.2}}
\psfrag{0.8}[t][t]{{\tiny 0.8}}
\psfrag{1.0}[t][t]{{\tiny 1.0}}
\psfrag{1.5}[t][t]{{\tiny 1.5}}
\psfrag{log(d)}[t][t]{ {\small $\log(d)$}}
\psfrag{log(r)}[t][t]{{\small $\log(r)$}}
\psfrag{MAP estimates}[t][t]{MAP estimates}
\includegraphics[height=\textwidth,  angle=-90]{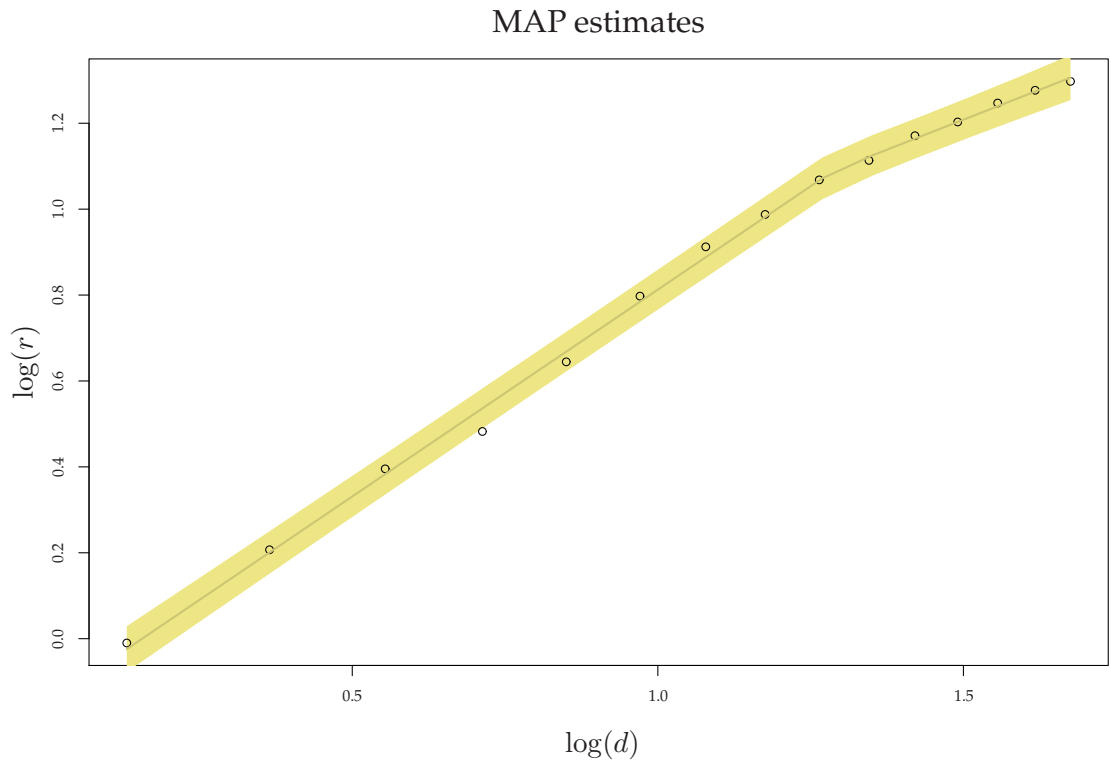}
\caption{Fitted curves using MAP for data from Dimini (with 95\% pointwise prediction interval). }
\label{fig:dimini}
\end{center}\end{figure}

Figure \ref{fig:dimini} shows the fitted curve using posterior modal estimates with a single
change point found to be around $\gamma_1=1.29$. \shortciteN{fan+b00} found
that the model with the highest posterior model probability contained one change point, with
mean 1.32 on the log scale. Similarly, our MLE estimates for the remaining parameters are
$\log(a_1)=-0.15 (-0.14)$, $\log(a_2)=0.37 (0.41)$, $b_1=0.96 (0.95)$ and
$b_2=0.56 (0.54)$, with the posterior mean estimate in parentheses quoted from  \shortciteN{fan+b00} for comparison. Finally, in terms of computation, the sampler used by \shortciteN{fan+b00} required 15,000,000 MCMC iterations, as the continuity constraint 
posed a problem for mixing. See also \shortciteN{sisson+fan07} for related discussion on mixing.

\subsection{Modelling extreme rainfall}\label{sec:rainfall}

We now consider the modelling of extreme levels of a sequence $\{X_t\},  t=1,\ldots T$, of daily rainfall measurements.
Following standard arguments from extreme value theory (e.g. \citeNP{coles01})
for a large enough threshold, $u$, the 
distribution of threshold exceedances, $Y_t=X_t-u$, conditional upon $X_t>u$, approximately follows a generalised Pareto distribution
\begin{equation}
\label{eqn:gpd}
	H(Y_t\leq y_t) = 1-\left(1+\frac{\xi_t}{\sigma_t}y_t\right)^{-1/\xi_t}
\end{equation}
defined on $\{y_t: y_t>0\mbox{ and }(1+\xi_t y_t/\sigma_t)>0\}$. Time-dependent parameters $\sigma_t$ and $\xi_t$ respectively determine scale and shape (through the rate of tail decay).

\begin{figure}[tbh]\begin{center}
\psfrag{Rainfall from Maiquetis, Venezuela}[t][t]{{\small Rainfall from Maiquetis, Venezuela}}
\psfrag{Rainfall (mm)}[b][b]{{\small Rainfall (mm)}}
\psfrag{Day of Year}[l][l]{{\small Day of year}}
\psfrag{0}[t][t]{{\tiny 0}}
\psfrag{100}[t][t]{{\tiny 100}}
\psfrag{200}[t][t]{{\tiny 200}}
\psfrag{300}[t][t]{{\tiny 300}}
\psfrag{50}[t][t]{{\tiny 50}}
\psfrag{150}[t][t]{{\tiny 150}}
\psfrag{250}[t][t]{{\tiny 250}}
\psfrag{350}[t][t]{{\tiny 350}}
\includegraphics[height=14cm,  angle=-90]{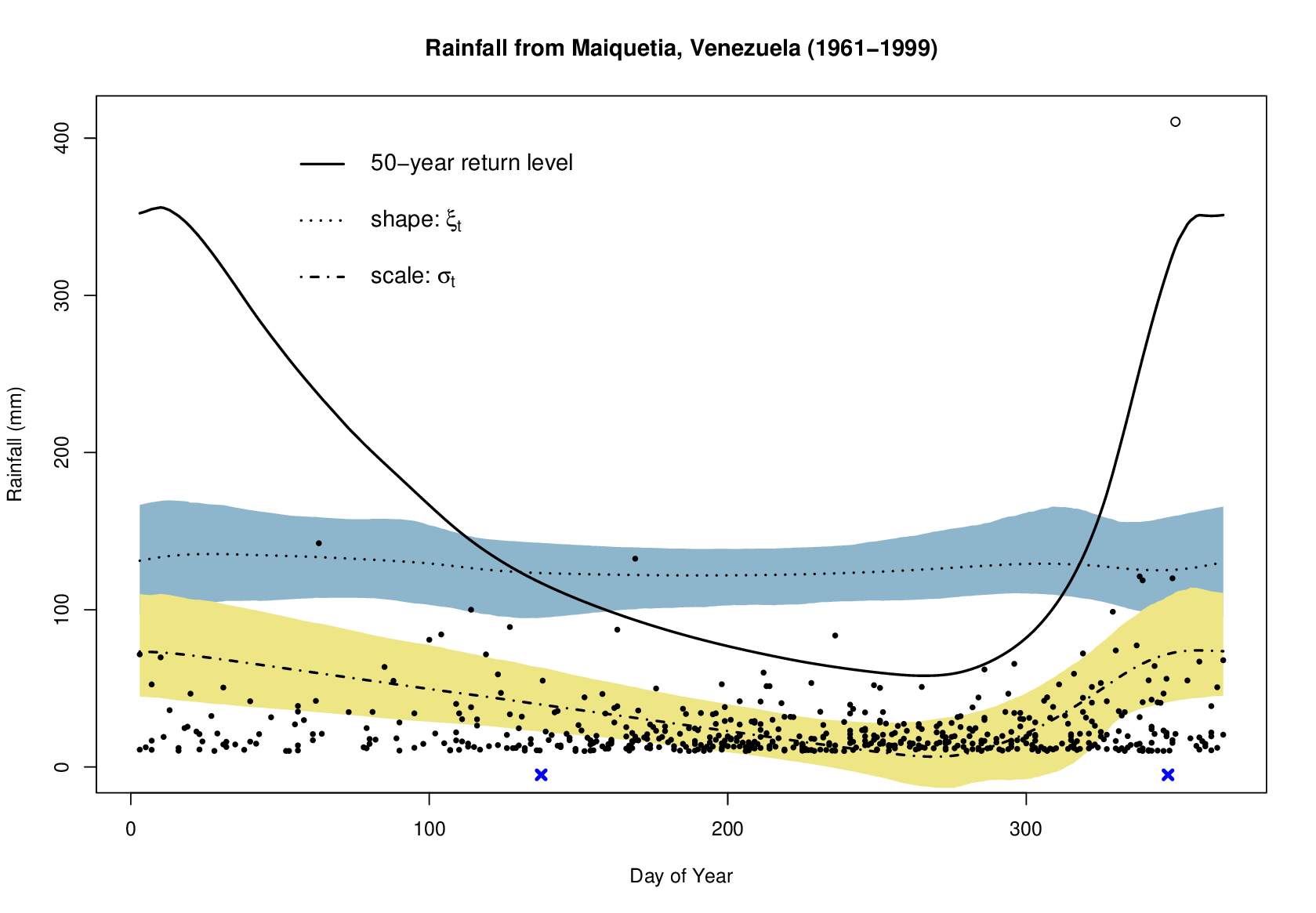}
\caption{\small{Daily rainfall measurements exceeding $u=10\,$mm recorded at Maiquetia International Airport, Venezuela during 1961--1999. Open circle indicates December 1999 event (not used to fit model). Curves denote (pointwise) posterior means and 95\% credibility regions for $\sigma_t$ and $\xi_t$, and 50-year return levels (curves linearly scaled for visualisation purposes). Crosses denote  MAP change-point estimates for two-season model under analysis of Sisson et al. (2006).}}
\label{fig:pareto}
\end{center}\end{figure}

We consider extreme daily
rainfall levels
recorded at Maiquetia International Airport, Venezuela, for the period 1961--1999. 
Particular interest in this series arises through an event in December 1999 which was almost three times greater than any previously recorded 
rainfall (Figure \ref{fig:pareto}, open circle).  
In a previous analysis, \shortciteN{sisson+pc06} modelled within-year seasonal variations using constant scale and shape parameters between seasonal change-points, where the number and location of change-points was  unknown. Accurate modelling of seasonal variability was demonstrated to be crucial in terms of making realistic predictions concerning the December 1999 event. The analysis of \shortciteN{sisson+pc06} implemented reversible jump MCMC with split/merge steps for between-model transitions. Between-model chain mixing was generally poor, necessitating long chain runs to ensure accurate posterior inference.

Here we model within-year variations by expressing both $\sigma_t$ and $\xi_t$ as  first-order curves $f(x)$ (Equation (\ref{eqn:gen}) with $P=1$) ,where $t=1,\ldots,366$ now specifically denotes the day of the year. The (unknown) location and number of knot points correspond to variations in the underlying seasonal climate.
As any temporal fluctuations in the distribution of rainfall extremes can reasonably be expected to affect both location and scale parameters simultaneously, we express both $\sigma_t$ and $\xi_t$ as functions of the same $\gamma$ and $z$ variables, but allow different $\beta_{z,\gamma}$ coefficients.
Given that the last day in the year is temporally adjacent to the first day of the following year, the model requires curve continuity at the yearly end points. This is  achieved by imposing the constraints $\sigma_0=\sigma_{366}$ and the first derivatives $\frac{d\sigma_t}{dt}\left|_{t=0}\right.=\frac{d\sigma_t}{dt}\left|_{t=366}\right.$ (and similarly for $\xi_t$). Specifically, this amounts to
\[
	\alpha_1=-\sum_{k=1}^{K}\eta_k(1-\gamma_k/366)
	\qquad\mbox{and}\qquad
	\eta_K=-\sum_{k=1}^{K-1}\eta_k
\]
for both $\sigma_t$ and $\xi_t$ (the indexing  of $\beta_{z,\gamma}$ coefficients on $\sigma_t$ and $\xi_t$ is suppressed for clarity).

The non-Gaussianity of the model means we are unable to analytically integrate out the $\beta_{z,\gamma}$ coefficients, and so we implement the algorithm in Section \ref{sec:non-gauss}.
In all, 5,000 MCMC iterations were obtained following 1,000 iterations burnin, for each iteration
we perform 10 updates of $\gamma$, and use $\kappa=10$ each update of $z$. 
Here, as maximum likelihood estimates of $\beta_{z,\gamma}$ under the generalised Pareto distribution require numerical optimisation of the likelihood, we modified the MLE estimate to be the maximum a posteriori estimates for improved sampler efficiency for the same computational effort. The covariance matrix scaling factor for the $\beta_{z,\gamma}$ proposal updates was set to $\delta_z=1$, $\delta_{\beta}=1/10$.
Prior specification was $\lambda=1$, $c=n$, $L=10$, and 10 equally spaced intervals over the range 1 to 366 were used.

Figure \ref{fig:pareto} displays the rainfall measurements plotted against the day of the year, with pointwise posterior means and 95\% credibility intervals for shape and scale parameters (scaled linearly for visualisation purposes). Also shown is the pointwise posterior predictive mean 50-year return level, defined as the rainfall level that is exceeded on average once every 50 years.
Following from (\ref{eqn:gpd}) this may be obtained as the value $z_{50}$ that is the solution of
\[
	\zeta_u\left(1+\frac{\xi_t}{\sigma_t}z_{50}\right)^{-1/\xi_t}=\frac{1}{50n_y}
\]
where $n_y=365.25$ is the average number of observations per year and $\zeta_u=\mbox{Pr}(X_t>u)$ is the probability that an individual observation exceeds the threshold, $u$.

The low return level around the middle of the year (in the ``wet'' season) corresponds to relatively low shape and scale parameters for this period, while conversely the high return level (in the ``dry'' season) corresponds to relatively high $\sigma_t$ and $\xi_t$. The timing of changes in the tail behaviour of the fitted Pareto density (as evidenced by variations in the 50-year return level) corresponds well with  previously identified MAP changepoints (indicated by $\times$'s in Figure \ref{fig:pareto}) for a two-seasonal model \shortcite{sisson+pc06}. The computation required for this inference was considerably less than for the earlier analysis.

\section{Discussion}

\label{sec5}
This article focuses on the auxiliary variable approach to the fitting of curves. This approach allows us to compute for the unknown number and location of the knots, via a Metropolis-within-Gibbs sampler. We have adopted the use of a spline regression model of the form (\ref{eqn:spline}). However, more sophisticated expressions can be found for curves (see for example \shortciteNP{denison+MS98}), to which the methods described here easily extend.

Our method depends, to some extent, on the specification of the intervals $I_k, k=1,\ldots K_{max}$ in which knots $\gamma$ may be found. The advantage of our approach over
\shortciteN{smith+kohn96} is that it gives a more accurate inference, particularly for small data sets. For instance in Example \ref{sec:tombs}, the MAP change-point is found to be between two data points while the method of  \shortciteN{smith+kohn96}  does not allow for this location. In all the examples presented in this paper we have only used non-overlapping intervals.  However, it is possible to allow overlapping intervals using, for example, an ordering constraint on the values of $\gamma$. We have also shown via simulated data sets that our method compares well with the method of  \shortciteN{dimatteo+gk01} which uses the reversible jump approach.

We have also provided a new Metropolis-within-Gibbs sampler algorithm to fit the regression model when the Gaussian error assumption is relaxed. In this case our sampler needs to include an additional step for the computation of the spline coefficients.  In particular, we advocate the use of MLEs in the construction of a proposal distribution for the coefficients when moving to a new model. 

Finally, we revisited two real examples of Bayesian change point detection, and showed that these types of problems may be converted to the variable selection setting, hence making use of the auxiliary variable approach. Many Bayesian change point analyses with unknown number and location of change points are computed with the use of complex implementations of the reversible jump algorithm (for example involving, split/merge and birth/death moves), as were the cases for the original analyses in Section \ref{sec4}. 
Although the reversible jump samplers can handle more complex, non-standard problems, we have found that our approach here is far simpler to implement. We were able to use standard statistical software R (\shortciteNP{venablesr05}) to implement both examples very efficiently.

\section{Supplemental materials}
The following supplemental materials are made available online.
\begin{description}
\item[Data and Computer Code] 
R programs to run the algorithms described in this article.
All simulated data sets and real data sets used in the examples are also included. Please
refer to the README files in the relevant directories for instructions. (curves.tar.zip, tarred zip file)

\end{description}
\section*{Acknowledgments}

The authors would like thank David Nott for useful discussion. YF and SAS are supported by the Australian Research Council through the
Discovery Project scheme (DP0877432). 

\bibliographystyle{chicago}

\bibliography{bayes}

\end{document}